\documentclass[review,12pt]{elsarticle}
\usepackage{color}
\usepackage{lineno,hyperref}
\modulolinenumbers[5]
\usepackage{amsmath,amsfonts,graphicx,epsfig}
\journal{Journal of \LaTeX\ Templates}
\newcommand{\ee}{\end{eqnarray}}
\newcommand{\be}{\begin{eqnarray}}
\newcommand*{\D}{{\rm d}}

\newcommand{\Lag}{{\cal L}}

\def\widebar{\overline}

\begin{document}

\begin{frontmatter}

\title{\bf Tracking our Universe to de Sitter by a Horndeski scalar}

\author{Cristiano Germani~\footnote{\tt email: germani@icc.ub.edu}}
\address{Institut de Ciencies del Cosmos (ICCUB), Universitat de Barcelona, Mart\`i i Franqu\'es 1,\\
E08028 Barcelona, Spain}
\author{Prado Mart\'in-Moruno~\footnote{\tt email: pradomm@ucm.es}}
\address{Departamento de F\'isica Te\'orica I, Universidad Complutense de Madrid, E-28040 Madrid, Spain}

\begin{abstract}

Assuming both that our Universe is evolving into a de Sitter space and a vanishing cosmological constant, leaves only the option that the observed acceleration is provided by a ``kinetic" energy of a scalar field. From an effective field theory point of view, the absence of Ostrogradsky instabilities restricts the choice to shift-symmetric Horndeski theories. Within these theories, we find the conditions for the existence of a de Sitter critical point  in a universe filled by matter, radiation and a Horndeski scalar. Moreover, we show that this point is a universal attractor and we provide the tracking trajectory. Therefore, if a de Sitter fixed point exists within these models, our Universe will eventually evolve into a de Sitter space. As an example, we have discussed the case of the combined Galileon-Slotheon system, in which the Galileon is kinetically non-minimal coupled to the Einstein tensor. Interestingly, we have also found that the tracker trajectory of this system does not follow previous literature assumptions.
\end{abstract}


\end{frontmatter}



\section{Introduction} 

Every observation points out to an accelerated expansion of our Universe, which is very well fit by a constant energy density. It seems quite plausible, in order to avoid an even worse {\it ``why now"} problem, that the Universe will evolve into a de Sitter space. This ``why now" problem is related to the question of why this constant energy density dominates our Universe evolution exactly at the right moment, in order to allow structure formations and life \cite{weinberg}. In this paper we work within this prejudice, namely the expectation that the Universe should naturally follow some kind of tracking trajectory to a de Sitter space, no matter what initial conditions for the cosmological fluids are used.

Obviously, the easiest option would seem to introduce a cosmological constant. However, by our knowledge of quantum field theory, tadpoles, generated by zero momentum loops of massive standard model particles, lead to a larger energy density than the one necessary to fit the data \cite{martin}. 
Assuming a mechanism canceling zero momentum contributions to the semiclassical gravitational equations, one is left to the problem of providing an asymptotically constant energy density via kinematical contributions.
If, in addition, Ostrogradsky instabilities are avoided, one is then left to consider asymptotically shift-symmetric Horndeski theories. To simplify our analysis, we will here only consider shift-symmetric models. One could in fact generalise our findings by considering an earlier potential contribution, however, we do not expect that this will change our conclusions significantly.
On the other hand, whether or not shift-symmetric Horndeski theories suffer from a similar quantum instability of the cosmological constant has to be yet proven in general. Nevertheless, encouraging results are already been found in reference \cite{initial}, where a sub-class of Horndeski theories (the covariant Galileons) with a de Sitter attractor, are found to be stable around their de Sitter fixed point. We will, however, not perform that analysis here and leave it for future work.

It is already well-known that covariant Galileons have a de Sitter attractor in the presence of dust matter and radiation whenever the parameters of the Lagrangian
satisfy particular relations \cite{DeFelice:2010pv} (see also reference \cite{Appleby:2011aa}). 
So, the current cosmological phase of accelerated expansion would be the result of approaching that de Sitter critical point.

Galileon models are just a particular case of Horndeski theories \cite{Horndeski:1974wa}, the most general scalar-tensor theories with second order equations of motion. It would then be extremely interesting to know what kind of Horndeski theories include a {\it stable} de Sitter critical point in presence of other cosmological fluids.

This question, for a sub-class of shift-symmetric models, has been investigated in reference \cite{DeFelice:2011bh} by requiring the existence of a rather restrictive form of the tracking trajectory to a stable de Sitter fixed point. On the other side, the conditions of whether a self-tuned de Sitter fixed point exists (whether stable or not), in the presence of any generic cosmological fluid, has been investigated in reference \cite{Martin-Moruno:2015bda}.

In this paper, we will focus on our Universe and consider a generic shift-invariant Horndeski dark energy scalar, matter and radiation. We will search for the conditions such that a future de Sitter point, whenever matter and radiation are diluted away, exists. The tracking trajectory to the de Sitter point is also found explicitly showing that the requirements of reference \cite{DeFelice:2011bh}, even within their selected theories, were too restrictive. Finally, whenever the de Sitter point exists, we have proven that it is stable. 


\section{Shift-symmetric Horndeski models on a spatially flat FRW}\label{sec:SS}

The Horndeski action is usually presented in two forms:

\begin{itemize}
\item\noindent {\it Original Horndeski form \cite{Horndeski:1974wa,Kobayashi:2011nu}}
\begin{eqnarray}\label{LH}
 \mathcal{L}_H&=&\delta^{\alpha\beta\gamma}_{\mu\nu\sigma}\left[\kappa_1\left(\phi,\,X\right)\nabla^\mu\nabla_\alpha\phi \,R_{\beta\gamma}{}^{\nu\sigma}
 +\frac{2}{3}\kappa_{1,X}\left(\phi,\,X\right)\nabla^\mu\nabla_\alpha\phi\nabla^\nu\nabla_\beta\phi\nabla^\sigma\nabla_\gamma\phi\right.\nonumber\\
 &+&\left.\kappa_3\left(\phi,\,X\right)\nabla_\alpha\phi\nabla^\mu\phi\,R_{\beta\gamma}{}^{\nu\sigma}+2\kappa_{3,X}\left(\phi,\,X\right)\nabla_\alpha\phi\nabla^\mu\phi\nabla^\nu\nabla_\beta\phi\nabla^\sigma\nabla_\gamma\phi\right]\nonumber\\
 &+&\delta_{\mu\nu}^{\alpha\beta}\left[F\left(\phi,\,X\right)\,R_{\alpha\beta}{}^{\mu\nu}+2F_{,X}\left(\phi,\,X\right)\nabla^\mu\nabla_\alpha\phi \nabla^\nu\nabla_\beta\phi
 +2\kappa_8\left(\phi,\,X\right)\nabla_\alpha\phi\nabla^\mu\phi\nabla^\nu\nabla_\beta\phi\right]\nonumber\\
 &-&6\left[F_{,\phi}\left(\phi,\,X\right)-X\,\kappa_8\left(\phi,\,X\right)\right]\nabla_\mu\nabla^\mu\phi
 +\kappa_9\left(\phi,\,X\right),\label{hor}
\end{eqnarray}
where $ X=-\nabla_\mu\phi\nabla^\mu\phi/2$,
$\kappa_i\left(\phi,\,X\right)$ are arbitrary functions, and\footnote{We have absorbed an additional $W(\phi)$ function in $F(\phi,\,X)$ \cite{Kobayashi:2011nu}.}
\begin{equation}\label{condF}
 F_{,X}=2\left(\kappa_3+2X\kappa_{3,X}-\kappa_{1,\phi}\right).
\end{equation}
\item{\it Modern form \cite{Kobayashi:2011nu}}
\be
\Lag_H&=&\sum_{i=2}^5 \Lag_i\ ,\cr
\Lag_2&=&K(\phi,X)\ ,\cr
\Lag_3&=&-G_3(\phi,X)\square\phi\ ,\cr
\Lag_4&=&G_4(\phi,X)R+G_{4X}(\phi,X)\left[(\square\phi)^2-\phi_{;\mu\nu}\phi^{;\mu\nu}\right]\ ,\cr
\Lag_5&=&G_5(\phi,X)G_{\mu\nu}\phi^{;\mu\nu}-\frac{1}{6}G_{5,X}\left[(\square\phi)^3+2\phi_{;\mu}{}^\nu\phi_{;\nu}{}^\alpha\phi_{;\alpha}{}^\mu-3\phi_{;\mu\nu}\phi^{;\mu\nu}\square\phi\right]\ .\label{modern}
\ee
\end{itemize}
The translation between Lagrangians \eqref{hor} and \eqref{modern} was first presented in reference \cite{Kobayashi:2011nu}. This is
\begin{eqnarray}
K&=&\kappa_9+4X\int^X\D X'\left(\kappa_{8\phi}-2\kappa_{3\phi\phi}\right),
\\
G_3&=&
6F_\phi-2X\kappa_8-8X\kappa_{3\phi}+2\int^X\D X'(\kappa_8-2\kappa_{3\phi}),
\\
G_4&=&2F-4X\kappa_3,
\\
G_5&=&-4\kappa_1\ .\label{dictionary}
\end{eqnarray}

Although \eqref{hor} looks way more cumbersome than \eqref{modern}, it turns out that the original Horndeski form greatly simplifies the background analysis on a spatially flat Friedman-Robertson-Walker metric
\be\label{FLRW}
ds^2=-dt^2+a(t)^2 d\vec{x}\cdot d\vec{x}\ ,
\ee
which is the background we aim to study here.
Thanks to the symmetries of the background, it is enough to consider the point-like Lagrangian defined in the minisuperspace $\{a,\,\phi\}$, where $\phi$ is homogenous.
Once the dependence on higher derivatives is integrated by parts, the point-like Lagrangian obtained from Lagrangian (\ref{LH}) takes the simple form \cite{Charmousis:2011ea}
\begin{equation}\label{Lsimple}
 L_{\rm H}\left(\phi,\,\dot\phi,\,a,\,\dot a\right)=a^3\sum_{i=0..3}X_i\left(\phi,\,\dot\phi\right)\,H^i,\qquad {\rm with}\qquad
 L_{\rm H}=\mathcal{V}_{(3)}^{-1}\int {\rm d}^3 x\,\mathcal{L}_H,
\end{equation}
where $\mathcal{V}_{(3)}$ is the spatial $3$-volume element,
$H=\dot a/a$ is the Hubble parameter, and an over-dot represents a derivative with respect to the cosmic time $t$. The functions $X_i$ are given by \cite{Charmousis:2011ea}
\begin{eqnarray}
 X_0&=&-\widebar Q_{7,\phi}\dot\phi+\kappa_9,\label{X0}\\
 X_1&=&-3\,\widebar Q_7+Q_7\dot\phi,\label{X1}\\
 X_2&=&12\,F_{,X}X-12\,F,\label{X2}\\
 X_3&=&-4\,\kappa_{1,X}\,\dot\phi^3,\label{X3}
\end{eqnarray}
with
\begin{eqnarray}
Q_7&=&\widebar Q_{7,\dot\phi}=6\,F_{,\phi}-3\,\dot\phi^2\kappa_8.\label{Q7}
\end{eqnarray}
Note that the Einstein--Hilbert term is contained in the Horndeski Lagrangian. We chose however to explicitly extract it from the Lagrangian \eqref{hor}. 
In this case, considering also the presence of other fluids, we define the new minisuperspace Lagrangian
\begin{equation}\label{Ltot}
 L=L_{\rm EH}+ L_{\rm H}+ L_{\rm f},
\end{equation}
where $L_{\rm f}$ accounts for minimally coupled perfect fluids.
With the Lagrangian \eqref{Ltot} it is easy to obtain dynamical equations. 

The modified Friedmann equation can be obtained by imposing 
$\mathcal{H}=\mathcal{H}_{\rm EH}+\mathcal{H}_{\rm H}+\mathcal{H}_{\rm f}=0$, where, as usual, the Hamiltonian ${\cal H}$ is the Legendre transformation of the Lagrangian $\cal L$. We then obtain the Friedmann equation
\begin{equation}\label{Friedmann}
 -3M_{p}^2H^2+\sum_{i=0..3}\left[(i-1)X_i+X_{i,\dot\phi}\dot\phi\right]H^i+\rho(a)=0\ ,
\end{equation}
where $\rho(a)$ is the conserved total energy density of the cosmic fluids and $M_p$ is the reduced Planck mass.
It is interesting to emphasize that the Friedmann equation does not contain second order derivatives of $\phi(t)$ or $a(t)$, as it was noted in reference \cite{Kobayashi:2011nu}. 

The field equation can then be obtained by considering the variation of the point-like Lagrangian (\ref{Lsimple}) with respect to the field $\phi$. This is
\begin{eqnarray}\label{field}
\sum_{i=0}^{3}   \left[X_{i,\phi}-3X_{i,\dot\phi}H- iX_{i,\dot\phi}\frac{\dot H}{H}-
 X_{i,\dot\phi \phi}\dot\phi-X_{i,\dot\phi\dot\phi}\ddot\phi\right] H^i=0.
\end{eqnarray}

We now assume, as discussed in the introduction, shift-invariance. This implies that by defining $\psi\equiv\dot\phi$ the functions appearing in equations \eqref{Friedmann} and \eqref{field} only depend on $\psi$.
The Friedmann equation (\ref{Friedmann}) can then be expressed as \cite{Martin-Moruno:2015kaa}
\begin{equation}\label{Friedshift}
 \Omega+\Omega_\psi=1,
\end{equation}
where
\begin{equation}\label{Omf}
 \Omega_\psi=\sum_{i=0}^{3}\left[(i-1)f_i(\psi)+\psi f_{i,\psi}(\psi)\right]h^{i-2},
\end{equation}
and we have defined the dimensionless quantity $h=H/H_{dS}$ and 
\begin{equation}\label{fs}
 f_i(\psi)=\frac{H_{dS}^{i-2}}{3M_{p}^2}X_i(\psi)\ .
\end{equation}
At the moment $H_{dS}$ is simply a mass scale, however, later on we will associate it to the Hubble constant at the de Sitter fixed point of the system.

By defining the number of e-foldings $N\equiv\ln a$, and denoting with a prime the derivatives with respect to $N$, the field equation (\ref{field}) can be re-written as \cite{Martin-Moruno:2015kaa}
\begin{equation}\label{Eq1}
 \psi'P_1\left(h,\,\psi\right)+h'P_2\left(h,\,\psi\right)+P_0\left(h,\,\psi\right)=0,
\end{equation}
with
\begin{eqnarray}
P_0\left(h,\,\psi\right)&=&3h\sum_{i=0}^{3}f_{i,\psi}(\psi)h^i,\label{P0}\\
P_1\left(h,\,\psi\right)&=&h\sum_{i=0}^{3} f_{i,\psi\psi}(\psi)h^i,\label{P1}\\
P_2\left(h,\,\psi\right)&=&\sum_{i=0}^{3}if_{i,\psi}(\psi)h^i.\label{P2}
\end{eqnarray}
Considering a universe filled with dust matter and radiation, we also have two additional equations to close the system. Those are
\begin{eqnarray}
\Omega_{\rm m}'&=&-\Omega_{\rm m}\left[3+2\frac{h'}{h}\right],\label{conm}\\
\Omega_{\rm r}'&=&-\Omega_{\rm r}\left[4+2\frac{h'}{h}\right],\label{conr}
\end{eqnarray}
with $\Omega$ given in equation (\ref{Friedshift}) equal to $\Omega_{\rm m}+\Omega_{\rm r}$.  We do not integrate equations (\ref{conm}) and (\ref{conr}) for later convenience.

\section{de Sitter attractors}\label{sec:dS}

As we have shown in the previous section, we have $3$ differential equations (equations (\ref{Eq1}), (\ref{conm}) and (\ref{conr})) and a constraint (the Friedmann equation (\ref{Friedshift}))
for $4$ variables ($h$, $\psi$, $\Omega_{\rm m}$, and $\Omega_{\rm r}$). In order to get an autonomous closed system, we substitute the constraint (\ref{Friedshift}) in both sides of equation (\ref{conm}).
Taking then into account equation (\ref{conr}), we obtain
\begin{equation}\label{Eq2}
 \psi'R_1\left(h,\,\psi\right)+h'R_2\left(h,\,\psi\right)+R_0\left(h,\,\psi,\,\Omega_{\rm r}\right)=0,
\end{equation}
with
\begin{eqnarray}
R_0\left(h,\,\psi,\,\Omega_{\rm r}\right)&=&-3(1-\Omega_\psi)-\Omega_{\rm r}\nonumber\\
&=&-3+3\sum_{i=0}^{3}\left[(i-1)f_i(\psi)+\psi f_{i,\psi}(\psi)\right]h^{i-2}-\Omega_{\rm r}\label{R0}\ ,\\
R_1\left(h,\,\psi\right)&=&\sum_{i=0}^3 \left[i f_{i,\psi}(\psi)+\psi f_{i,\psi\psi}(\psi)\right]h^{i-2}\label{R1}\ ,\\
R_2\left(h,\,\psi\right)&=&h^{-1}\left\{-2+\sum_{i=0}^3i\left[(i-1)f_i(\psi)+\psi f_{i,\psi}(\psi)\right]h^{i-2} \right\}.\label{R2}
\end{eqnarray}

Now, combining equations (\ref{Eq1}) and (\ref{Eq2}), we get
\begin{eqnarray}\label{dh}
 h'=\frac{R_0\left(h,\,\psi,\,\Omega_{\rm r}\right)P_1\left(h,\,\psi\right)-P_0\left(h,\,\psi\right)R_1\left(h,\,\psi\right)}{P_2\left(h,\,\psi\right)R_1\left(h,\,\psi\right)-R_2\left(h,\,\psi\right)P_1\left(h,\,\psi\right)},
\end{eqnarray}
and
\begin{eqnarray}\label{dpsi}
 \psi'=\frac{R_2\left(h,\,\psi\right)P_0\left(h,\,\psi\right)-P_2\left(h,\,\psi\right)R_0\left(h,\,\psi,\,\Omega_{\rm r}\right)}{P_2\left(h,\,\psi\right)R_1\left(h,\,\psi\right)-R_2\left(h,\,\psi\right)P_1\left(h,\,\psi\right)}.
\end{eqnarray}
Substituting equation (\ref{dh}) in equation (\ref{conr}), we obtain 
\begin{eqnarray}
\Omega_{\rm r}'&=&-\Omega_{\rm r}\left[4+\frac{2}{h}\frac{R_0\left(h,\,\psi,\,\Omega_{\rm r}\right)P_1\left(h,\,\psi\right)-P_0\left(h,\,\psi\right)R_1\left(h,\,\psi\right)}{P_2\left(h,\,\psi\right)R_1\left(h,\,\psi\right)-R_2\left(h,\,\psi\right)P_1\left(h,\,\psi\right)}\right].\label{dO}
\end{eqnarray}
Equations (\ref{dh}), (\ref{dpsi}) and (\ref{dO}) form an autonomous closed system suitable for analysing the existence of critical points.

\subsection{de Sitter critical point in Horndeski and tracker solution}\label{sec:cpt}
We now look for the existence of a de Sitter critical point characterised by a Hubble scale $H_{dS}$, i.e. a point in which 
\begin{equation}\label{cpdS}
 h_{dS}=1,\qquad\Omega_{\rm r,dS}=0,\qquad {\rm and}\qquad \psi_{dS}\,\,\,{\rm such\,\,\,that}\,\,\,\Omega_\psi(h=1,\psi_{dS})=1.
\end{equation}
Taking into account equation (\ref{Omf}), the last condition implies
\begin{equation}\label{c1}
 \sum_{i=0}^{3}\left[(i-1)f_i(\psi_{dS})+\psi_{dS} f_{i,\psi}(\psi_{dS})\right]=1.
\end{equation}
Note that the conditions (\ref{cpdS}) directly imply that $\Omega_{\rm r}'=0$ in equation (\ref{dO}).
As $R_0\left(h_{dS},\,\psi_{dS},\,\Omega_{\rm r,dS}\right)=0$, imposing $h'=0$ and $\psi'=0$ it is easy to convince ourselves that the only solution is $P_0\left(h_{dS},\,\psi_{dS}\right)=0$, as can be seen by combining \eqref{dh} and \eqref{dpsi}. 

Thus, we have
\begin{equation}\label{ca2}
\sum_{i=0}^{3}f_{i,\psi}(\psi_{dS})=0\ .
\end{equation}
This condition can be used to simplify the earlier condition (\ref{c1}) into\begin{equation}\label{ca1}
\sum_{i=0}^{3}(i-1)f_{i}(\psi_{dS})=1.
\end{equation}
Therefore, a Universe filled by dust, radiation and a shift-symmetric Horndeski scalar has a de Sitter fixed point if and only if there exists a real solution $\psi_{dS}$ to equations (\ref{ca2}) and (\ref{ca1}). This de Sitter point is characterised by $H_{dS}$.

Furthermore, one can now calculate the Jacobian matrix of the system given by equations (\ref{dh}), (\ref{dpsi}) and (\ref{dO}) and evaluate it at the critical point \eqref{cpdS}, satisfying equations (\ref{ca2}) and (\ref{ca1}). The eigenvalues of this matrix are then
\begin{equation}
 \lambda_1=-4,\qquad\lambda_2=-3,\qquad\lambda_3=-3\ ,
\end{equation}
independently upon the explicit form of $f_i$'s. This implies that, whenever the de Sitter critical point exists, it is a universal attractor for {\it any} shift-invariant Horndeski models. This result is in agreement and generalise the one of reference \cite{Martin-Moruno:2015kaa}.

On the other hand, we can see this universal stability from a slightly different perspective. By using the modern formalism, all shift invariant Horndeski models can be written in terms of a conserved current $J$ \cite{Kobayashi:2011nu}
\be\label{cons}
\frac{d}{dt}\left(a^3 J\right)=0\ ,
\ee
if and only if (shift invariance condition)
\begin{eqnarray}
P_\phi &\equiv&
K_\phi-2X\left(G_{3\phi\phi}+\ddot\phi G_{3\phi X}\right)
+6\left(2H^2+\dot H\right)G_{4\phi}
+6H\left(\dot X+2HX\right)G_{4\phi X}
\nonumber\\&&
-6H^2XG_{5\phi\phi}+2H^3X\dot\phi \,G_{5\phi X}=0\ .
\end{eqnarray}
where
\begin{eqnarray}\label{J}
J&=&\dot\phi \,K_X+6HXG_{3X}-2\dot\phi G_{3\phi}
+6H^2\dot\phi\left(G_{4X}+2XG_{4XX}\right)-12HXG_{4\phi X}
\nonumber\\&&
+2H^3X\left(3G_{5X}+2XG_{5XX}\right)
-6H^2\dot\phi\left(G_{5\phi}+XG_{5\phi X}\right)\ . 
\end{eqnarray}
There are possibly two non-trivial solutions of equation \eqref{cons}: $J=0$ and $J=\frac{J_0}{a^3}$, where $J_0$ is a constant. The trajectory in the phase space satisfying $J(\psi,\,H)=0$ is obviously an attractor in any expanding Universe. In particular, if a non-trivial solution for $\psi(H)$ of $J\left(\psi, H\right)=0$ exists, no matter what the initial conditions are, $\psi$ will asymptotically (in time) tend to that solution. We call this solution the tracker and we will denote it as $\psi_{\rm tracker}(H)$. 

The conserved current associated with the shift symmetry could also be obtained directly from the minisuperspace Lagrangian (\ref{Lsimple}). So, we could have defined 
\begin{equation}\label{miniJ}
J=a^{-3}\frac{ \partial L}{\partial \dot\phi}=\sum_{i=0}^3X_{i,\psi}H^i\ .
\end{equation}
Therefore the tracker condition $J=0$ is equivalent to \eqref{ca2} whenever both radiation and matter vanish. Thus, a universe with a scalar on the tracker trajectory necessarily evolves to the de Sitter critical point in the future. The reason is that the equation $J=0$ contains no explicit scale factor, on the contrary, conservation equations for  radiation and matter imply their decay in time with the scale factor. Then, if a trajectory of $\psi$ is chosen such to include a de Sitter critical point, it will always be reached by the Universe no matter what the initial conditions for matter and /or radiation are. 
This explains the universal stability found through the dynamical system analysis. 
\section{The Slotheonic Galileon model}\label{sec:app}

To warm up and provide a non-trivial check of our formalism against previous literature, we will start by analysing the pure covariant Galileon model considered in reference \cite{DeFelice:2011bh}.
Galileon models are a particular case of shift-symmetric Horndeski models with \cite{DeFelice:2011bh}
\begin{eqnarray}\label{Gg}
 K=-c_2X,\quad G_3=\frac{c_3}{M^3}X,\quad G_4=-\frac{c_4}{M^6}X^2,\quad G_5=\frac{3c_5}{M^9}X^2,
\end{eqnarray}
where $M$ is a mass scale, which is related to the de Sitter point by $M^3=M_{p}H_{dS}^2$.

Considering the dictionary between Lagrangians \eqref{hor} and \eqref{modern} given in \eqref{dictionary}, we have
\begin{eqnarray}
\kappa_1&=& -\frac{3\,c_5}{16M^9}X^2,\quad \kappa_3=\frac{c_4}{4M^6}X,\quad F=-\frac{3\,c_4}{8M^6}X^2,\\
\kappa_8&=&-\frac{c_3}{2M^3}\,\ln\left(-\frac{X}{2}\right),\quad \kappa_9= \frac{c_2}{2}X.
\end{eqnarray}
It is very interesting to note that the original Horndeski coefficients are not analytical in $X$ while the $G_i$'s are.

Now, taking into account equations (\ref{X0})-(\ref{X3}), we obtain the functions appearing in the minisuperspace Lagrangian (\ref{Lsimple}). These are
\begin{eqnarray}\label{Xgalileons}
X_0=-\frac{c_2}{2}\psi^2,\quad X_1=\frac{c_3}{M^3}\psi^3,\quad X_2=-\frac{9\,c_4}{2M^6}\psi^4,\quad X_3=\frac{3\,c_5}{M^9}\psi^5,
\end{eqnarray}
where we emphasize again that the contribution $-3M_{p}^2$ to the $X_2$ function has already being considered in Lagrangian (\ref{Ltot}).

\subsection{Critical point and tracker solution for Galileons}\label{sec:G}
We now consider the conditions for a covariant Galileon Lagrangian to have a de Sitter critical point. 
Taking into account the equations contained in expression (\ref{Xgalileons}), in equations (\ref{ca2}) and (\ref{ca1}), we get
\begin{equation}\label{g1}
 -c_2\psi_{dS}+\frac{3\,c_3H_{dS}}{M^3}\psi_{dS}^2-\frac{18\,c_4H_{dS}^2}{M^6}\psi_{dS}^3+\frac{15\,c_5H_{dS}^3}{M^9}\psi_{dS}^4=0,
\end{equation}
and
\begin{equation}\label{g2}
 c_2\psi_{dS}^2-\frac{9\,c_4H_{dS}^2}{M^6}\psi_{dS}^4+\frac{12\,c_5H_{dS}^3}{M^9}\psi_{dS}^5=6M_{p}^2H_{dS}^2.
\end{equation}
Defining $x_{dS}=\psi_{dS}/(H_{dS}M_{p})$, $\alpha= c_4\,x_{dS}^4$, and $\beta= c_5\,x_{dS}^5$, according to reference \cite{DeFelice:2010pv}, equation (\ref{g1}) can be written as
\begin{equation}\label{cg1}
 c_2\,x_{dS}^2=6+9\alpha-12\beta.
\end{equation}
Combining equation (\ref{cg1}) with equation (\ref{g2}), we finally get
\begin{equation}\label{cg2}
 c_3\,x_{dS}^3=2+9\alpha-9\beta.
\end{equation}
Equations (\ref{cg1}) and (\ref{cg2}) are precisely the same expression obtained in reference \cite{DeFelice:2010pv} for the existence of a de Sitter critical point. Moreover,
given the analysis presented in the previous section, we already know that this critical point will be an attractor, as also found in reference \cite{DeFelice:2010pv}.
On the other hand, it should be noted that the region of allowed parameters in \eqref{cg1} and \eqref{cg2} will be further restricted by requiring classical \cite{DeFelice:2010pv} and quantum \cite{initial} stability.

Finally, the conserved current for the Galileon models is 
\be
J=\psi\left(a_0+a_1 y+a_2 y^2+a_3y^3\right)\ ,\label{Jg0}
\ee
where $y=\frac{\psi H}{M^3}$ and
\be
a_0=-c_2\ ,\quad
a_1=3 \,c_3\ ,\quad
a_2=-18\,c_4\ ,\quad
a_3=15\,c_5\ ,
\ee
which can be obtained using expressions (\ref{Gg}) in equation (\ref{J}) or expressions (\ref{Xgalileons}) in equation (\ref{miniJ}).
The tracker solution for the Galileon models is simply found by imposing $J=0$ in equation (\ref{Jg0}). Apart from the trivial solution $\psi=0$, we have $y_{\rm tr}={\rm constant}$, being the constant given by
\begin{equation}\label{tr}
a_0+a_1 y_{\rm tr}+a_2 y_{\rm tr}^2+a_3y_{\rm tr}^3=0,
\end{equation}
so, $\psi_{\rm tracker}=c/H$ is the Galileon tracker trajectory found in the literature \cite{DeFelice:2010pv}. Note that when the Hubble scale reaches  the value $H_{dS}$ in equation (\ref{tr}), the equation (\ref{g1}) is reproduced with $\psi_{\rm tracker}=\psi_{dS}$. Thus, we would like to stress once more that whenever there is a real solution for the system (\ref{g1})-(\ref{g2}), the de Sitter attractor exists and it is contained in the tracker trajectory.

Driven by similar results, the authors in reference \cite{DeFelice:2011bh} searched for tracker solutions within shift-symmetric Horndeski models with the restriction that the $K$s and $G$s functions are only single powers of $X$. However, the same authors also assumed a functional form of the type $H\psi^p={\rm const}$, where $p$ is a constant. Of course, by the above discussion, we already know that such constraint might be too strong and would miss entire classes of tracking solutions defined by $J=0$ but {\it not} of the form $H\psi^p={\rm const}$, as we shall see it in a specific example.

\subsection{Critical point and tracker solution for Slotheonic Galileons}\label{sec:S}

Now that we have tested our algorithm with the Galileon models, we can investigate a new case.
As a working example let us take a Galileon kinetically non-minimally coupled to the Einstein tensor. This coupling has been doubted slotheonic coupling in reference \cite{sloth} because it generically makes any scalar ``slower" by enhancing the gravitational friction \cite{yuki}. The slotheonic coupling is $\frac{1}{2 M_*^2}G^{\mu\nu}\partial_\mu\phi\partial_\nu\phi$, which, in the language of reference \cite{Kobayashi:2011nu}, is obtained by chosing $G_5=-\frac{\phi}{2 M_*^2}$ or, equivalently, $G_4=\frac{X}{2 M_*^2}$. Note that, in the case in which $c_4=0$, the Slotheonic Galileon is a sub-class of the theories studied in reference \cite{DeFelice:2011bh}.

Before discussing the existence of a de Sitter critical point, we would like to point out a striking difference in the tracker solution between the pure covariant Galileon and the Slotheonic Galileon. 
It is straightforward to see that the tracker condition $J=0$ now implies
\be\label{trackerG+}
a_0+a_1 y_{\rm tr}+a_2y_{\rm tr}^2+a_3 y_{\rm tr}^3+3 \frac{H^2}{M_*^2}=0\ ,
\ee
which greatly differs from the earlier literature hypothesis that the tracker solution of this system should be of the form $H\psi^p={\rm constant}$ \cite{DeFelice:2011bh}. Note that this is not an artefact of a bad choice of a frame (i.e. Jordan versus Einstein). The reason is that there exist no conformal transformation of the Slotheon/metric that leads either to the Einstein or to the Jordan frame (see e.g. \cite{sloth}).

The conditions for a de Sitter attractor are found by noticing that the Slotheon simply provides a shift of the $X_2$ function as follows
\be\label{shiftedX2}
X_2=\frac{3}{2}\psi^2\left(-\frac{3\,c_4}{M^6}\psi^2+\frac{1}{M_*^2}\right)\ .
\ee
Taking into account equations (\ref{ca2}) and (\ref{ca1}), we obtain the two new constraints 
\be\label{newregion}
c_2\,x_{dS}^2 &=& 6+9\alpha-12\beta-3\gamma\ ,\label{cs1}\\
c_3\,x_{dS}^3 &=& 2+9\alpha-9\beta-2\gamma\ , \label{cs2}
\ee
where we have defined $\gamma=H_{dS}^2/M_*^2$. 

As in the previous case, we will not further investigate the restriction of the parameter space \eqref{cs1} and \eqref{cs2} due to classical and quantum stability and leave this for future work. The reason is that the aim of this section is only to provide a non-trivial new example of a Horndeski theory with a de Sitter critical point and show a novel attractor behaviour.  

We want to emphasise that to avoid any hierarchy of scales, all $c_i\sim {\cal O}(1)$ and, in the covariant Galileon, $M^3= H_{dS}^2 M_p$  \cite{initial,DeFelice:2010pv}. On the same grounds one can show that $\psi_{dS}\sim M_p H_{dS}$ \cite{initial}. If we now again impose no hierarchies of scales also for the Slotheonic Galileon, we find $M_*\sim H_{dS}$, in particular we can define $M_*^2\equiv H_{dS}^2/c_*$ (or $\gamma=c_*$), with $c_*={\cal O}(1)$. This is indeed the right scale one could guess for a slotheonic interaction. The reason is the following: suppose we assume that $M$ is the unique suppression scale of the system, then the Slotheon operator expanded on a Minkowski background will look like $\frac{\partial^2 \bar h^{\alpha\beta}}{M_*^2 M_p}\partial_\alpha\phi\partial_\beta\phi$, where $\bar h$ is the canonicalised graviton. Then by fixing $M_*^2 M_p\sim M^3$ we readily obtain $M_*\sim H_{dS}$.

Finally, during radiation epoch the ratio $\frac{H^2}{M_*^2}\gg 1$, i.e. the Slotheon will be in the gravitationally enhanced friction regime \cite{yuki}. There, by taking into account that $c_i\sim {\cal O}(1)$, in order to solve equation \eqref{trackerG+}, one needs to have 
\be
\psi_i\gg \left(\frac{H_{dS}}{H_i}\right) H_{dS} M_p\ ,\label{psic}
\ee
where $i$ denotes the initial value during radiation. Then, if this condition is reached, one finds (forgetting order one factors)
\be\label{psis}
\psi_i\sim \left(\frac{H_i}{H_{dS}}\right)^{\frac{2}{n}-1} \psi_{dS}\ .
\ee
Finally, one finds the following hierarchies
\be
\psi_i&\gg& \psi_{dS}\ {\rm for}\ n=1\ {\rm (cubic\ Galileon\ domination)}\ ,\nonumber\cr
\psi_i&\sim& \psi_{dS}\ {\rm for}\ n=2\ {\rm (quartic\ Galileon\ domination)}\ ,\nonumber\cr
\psi_i&\ll& \psi_{dS}\ {\rm for}\ n=3\ {\rm (quintic\ Galileon\ domination)}\nonumber\ .
\ee

\section{Conclusions}\label{sec:c}

In the case in which a cosmological constant is absent, the observed Universe acceleration may be obtained by a scalar field ``kinetic energy". Forbidding ghosts instabilities, this leads us to the general class of shift-symmetric Horndeski scalar-tensor theories. The avoidance of a cosmological conspiracy, where the dark energy would only dominate now, implies that our Universe is approaching to a de Sitter point in the far future. Then, by focusing on our Universe that is filled by dust matter, radiation and a dark energy scalar we found the conditions for which a de Sitter point exists in the future. We then show, within the shift-symmetric Horndeski models, that a de Sitter point is a universal attractor and we thus provide the generic tracking trajectory to that point.

Finally, we have applied our general formalism to specific examples. As a consistency check, we have studied the Galileons sub-class of Horndeski theories and recovered the results already found in reference \cite{DeFelice:2010pv}. Then, we have extended the Galileon theory by introducing a Slotheonic interaction, namely the coupling of the Galileon field to the Einstein tensor, which again represents a sub-class of Horndeski theories. This example is very interesting as, in addition to contain a de Sitter stable fixed point, have a tracking trajectory that greatly differs from the one assumed in previous literature (see for example reference \cite{DeFelice:2011bh}).

\section*{Acknowledgements}
\noindent CG would like to thank Emilio Bellini for many useful discussions.
CG is supported by the Ramon y Cajal program and partially supported by the Unidad de Excelencia Mar\'ia de Maeztu Grant No. MDM-2014-0369 and FPA2013-46570-C2-2-P grant.
The work of PMM has been supported by the projects FIS2014-52837-P (Spanish MINECO) and FIS2016-78859-P (AEI/FEDER, UE).

\section*{References}

\bibliography{bibliography}

\begin{thebibliography}{10}
\expandafter\ifx\csname url\endcsname\relax
  \def\url#1{\texttt{#1}}\fi
\expandafter\ifx\csname urlprefix\endcsname\relax\def\urlprefix{URL }\fi
\expandafter\ifx\csname href\endcsname\relax
  \def\href#1#2{#2} \def\path#1{#1}\fi

\bibitem{weinberg}
S.~Weinberg, {Anthropic Bound on the Cosmological Constant}, Phys. Rev. Lett.
  59 (1987) 2607.
\newblock \href {http://dx.doi.org/10.1103/PhysRevLett.59.2607}
  {\path{doi:10.1103/PhysRevLett.59.2607}}.

\bibitem{martin}
J.~Martin, {Everything You Always Wanted To Know About The Cosmological
  Constant Problem (But Were Afraid To Ask)}, Comptes Rendus Physique 13 (2012)
  566--665.
\newblock \href {http://arxiv.org/abs/1205.3365} {\path{arXiv:1205.3365}},
  \href {http://dx.doi.org/10.1016/j.crhy.2012.04.008}
  {\path{doi:10.1016/j.crhy.2012.04.008}}.

\bibitem{initial}
C.~Germani, {Initial conditions for the Galileon dark energy}, Phys. Dark Univ.
  15 (2017) 1--6.
\newblock \href {http://arxiv.org/abs/1609.06598} {\path{arXiv:1609.06598}},
  \href {http://dx.doi.org/10.1016/j.dark.2016.11.003}
  {\path{doi:10.1016/j.dark.2016.11.003}}.

\bibitem{DeFelice:2010pv}
A.~De~Felice, S.~Tsujikawa, {Cosmology of a covariant Galileon field}, Phys.
  Rev. Lett. 105 (2010) 111301.
\newblock \href {http://arxiv.org/abs/1007.2700} {\path{arXiv:1007.2700}},
  \href {http://dx.doi.org/10.1103/PhysRevLett.105.111301}
  {\path{doi:10.1103/PhysRevLett.105.111301}}.

\bibitem{Appleby:2011aa}
S.~Appleby, E.~V. Linder, {The Paths of Gravity in Galileon Cosmology}, JCAP
  1203 (2012) 043.
\newblock \href {http://arxiv.org/abs/1112.1981} {\path{arXiv:1112.1981}},
  \href {http://dx.doi.org/10.1088/1475-7516/2012/03/043}
  {\path{doi:10.1088/1475-7516/2012/03/043}}.

\bibitem{Horndeski:1974wa}
G.~W. Horndeski, {Second-order scalar-tensor field equations in a
  four-dimensional space}, Int. J. Theor. Phys. 10 (1974) 363--384.
\newblock \href {http://dx.doi.org/10.1007/BF01807638}
  {\path{doi:10.1007/BF01807638}}.

\bibitem{DeFelice:2011bh}
A.~De~Felice, S.~Tsujikawa, {Conditions for the cosmological viability of the
  most general scalar-tensor theories and their applications to extended
  Galileon dark energy models}, JCAP 1202 (2012) 007.
\newblock \href {http://arxiv.org/abs/1110.3878} {\path{arXiv:1110.3878}},
  \href {http://dx.doi.org/10.1088/1475-7516/2012/02/007}
  {\path{doi:10.1088/1475-7516/2012/02/007}}.

\bibitem{Martin-Moruno:2015bda}
P.~Martin-Moruno, N.~J. Nunes, F.~S.~N. Lobo, {Horndeski theories self-tuning
  to a de Sitter vacuum}, Phys. Rev. D91~(8) (2015) 084029.
\newblock \href {http://arxiv.org/abs/1502.03236} {\path{arXiv:1502.03236}},
  \href {http://dx.doi.org/10.1103/PhysRevD.91.084029}
  {\path{doi:10.1103/PhysRevD.91.084029}}.

\bibitem{Kobayashi:2011nu}
T.~Kobayashi, M.~Yamaguchi, J.~Yokoyama, {Generalized G-inflation: Inflation
  with the most general second-order field equations}, Prog. Theor. Phys. 126
  (2011) 511--529.
\newblock \href {http://arxiv.org/abs/1105.5723} {\path{arXiv:1105.5723}},
  \href {http://dx.doi.org/10.1143/PTP.126.511}
  {\path{doi:10.1143/PTP.126.511}}.

\bibitem{Charmousis:2011ea}
C.~Charmousis, E.~J. Copeland, A.~Padilla, P.~M. Saffin, {Self-tuning and the
  derivation of a class of scalar-tensor theories}, Phys. Rev. D85 (2012)
  104040.
\newblock \href {http://arxiv.org/abs/1112.4866} {\path{arXiv:1112.4866}},
  \href {http://dx.doi.org/10.1103/PhysRevD.85.104040}
  {\path{doi:10.1103/PhysRevD.85.104040}}.

\bibitem{Martin-Moruno:2015kaa}
P.~Martin-Moruno, N.~J. Nunes, {Attracted to de Sitter II: cosmology of the
  shift-symmetric Horndeski models}, JCAP 1509~(09) (2015) 056.
\newblock \href {http://arxiv.org/abs/1506.02497} {\path{arXiv:1506.02497}},
  \href {http://dx.doi.org/10.1088/1475-7516/2015/09/056}
  {\path{doi:10.1088/1475-7516/2015/09/056}}.

\bibitem{sloth}
C.~Germani, L.~Martucci, P.~Moyassari, {Introducing the Slotheon: a slow
  Galileon scalar field in curved space-time}, Phys. Rev. D85 (2012) 103501.
\newblock \href {http://arxiv.org/abs/1108.1406} {\path{arXiv:1108.1406}},
  \href {http://dx.doi.org/10.1103/PhysRevD.85.103501}
  {\path{doi:10.1103/PhysRevD.85.103501}}.

\bibitem{yuki}
C.~Germani, Y.~Watanabe, {UV-protected (Natural) Inflation: Primordial
  Fluctuations and non-Gaussian Features}, JCAP 1107 (2011) 031, [Addendum:
  JCAP1107,A01(2011)].
\newblock \href {http://arxiv.org/abs/1106.0502} {\path{arXiv:1106.0502}},
  \href {http://dx.doi.org/10.1088/1475-7516/2011/07/031,
  10.1088/1475-7516/2011/07/A01} {\path{doi:10.1088/1475-7516/2011/07/031,
  10.1088/1475-7516/2011/07/A01}}.

\end{thebibliography}

\end{document}